\documentstyle[prc,aps,epsfig,multicol]{revtex}
\draft
\begin{document}
\title{Charged particle rapidity distributions at RHIC}
\author{Zi-wei Lin$^1$, Subrata Pal$^1$, C.M. Ko$^1$, Bao-An Li$^2$, 
and Bin Zhang$^2$}
\address{$^1$Cyclotron Institute and Physics Department,
Texas A\&M University, College Station, Texas 77843-3366}
\address{$^2$Department of Chemistry and Physics, Arkansas State University,
P.O. Box 419, State University, Arkansas 72467-0419}

\maketitle

\begin{abstract}
Using a multiphase transport model (AMPT), which includes both 
initial partonic and final hadronic interactions, we study the 
rapidity distributions of charged particles such as protons, 
antiprotons, pions, and kaons in heavy ion collisions at RHIC. 
The theoretical results for the total charged particle multiplicity 
at midrapidity are consistent with those measured
by the PHOBOS collaboration in central Au+Au collisions at 
$\sqrt s$ = 56 and 130 AGeV. We find that these hadronic observables 
are much more sensitive to the hadronic interactions than the partonic 
interactions.

\end{abstract}

\vspace{0.5cm}
\vspace{0.5cm}

\begin{multicols}{2}

Collisions of nuclei at high energies offer the possibility to 
subject nuclear matter to the extreme conditions of large compression
and high excitation energies. Studies based on both non-equilibrium
transport models \cite{transport} and equilibrium thermal models \cite{thermal}
have shown that the experimental data from heavy ion collisions at 
SIS, AGS and SPS, where the center of mass collision energies are, 
respectively, about 3, 5 and 17 AGeV, are consistent with the formation 
of a hot dense nuclear matter in the initial stage of collisions. 
With the Relativistic Heavy Ion Collider (RHIC) at Brookhaven National 
Laboratory, which can reach a center of mass energy of 200 AGeV, 
the initial energy density is expected to exceed that for the transition
from the hadronic matter to the quark-gluon plasma. Experiments
at RHIC thus provide the opportunity to recreate the matter which
is believed to have existed during the first microsecond after the
Big Bang and to study its properties.

Recently, charged particle multiplicity near mid-rapidity has been
measured in central Au+Au collisions at $\sqrt s=$ 56 and 130 AGeV
at RHIC by the PHOBOS collaboration \cite{phobos}. The observed charged 
particle density per participant is found to be compatible with 
the predictions of the HIJING model that includes particle production 
from minijets produced in hard-scattering processes \cite{hijing}. 
Although the HIJING model implements the parton energy loss  
via jet quenching \cite{xnwang}, it does not include explicit interactions 
among minijet partons and the final-state interactions among hadrons. 
Other models have also been used to understand the data 
from the PHOBOS collaboration. The LEXUS model \cite{lexus}, 
which is based on a linear extrapolation of ultra-relativistic 
nucleon-nucleon scattering to nucleus-nucleus collisions,
predicts too many charged particles compared with the PHOBOS data 
\cite{jeon}. On the other hand, the hadronic cascade model LUCIFER 
\cite{lucifer} predicts a charged particle multiplicity near mid-rapidity 
that is comparable to the PHOBOS data \cite{kahana}. 
In this paper, we shall use a recently developed multiphase transport 
model (AMPT) \cite{ampt}, that includes both partonic and
hadronic interactions, to study their effects  
not only on the total charged particle multiplicity but also
on those of kaons, protons, and antiprotons.

In the AMPT model, the initial conditions are obtained from  
the HIJING model \cite{hijing} by using a Woods-Saxon radial shape for the   
colliding nuclei and including the nuclear shadowing effect on parton
production via the gluon recombination mechanism of Mueller-Qiu 
\cite{amueller}. After the colliding nuclei pass through each other, 
the Gyulassy-Wang model \cite{gyulassy} is then used to generate the 
initial space-time information of partons. In the default HIJING, 
these minijet partons are allowed to lose energy via the gluon 
splitting mechanism and transfer their energies to the nearby strings 
associated with initial soft interactions. 
Such jet quenching is replaced in the AMPT model by explicitly taking into
account parton-parton collisions via Zhang's Parton Cascade (ZPC) 
\cite{zpc}. At present, only gluon elastic scatterings are included,
so the partons do not suffer any inelastic energy loss as they traverse 
the dense matter. After partons stop interacting, they combine 
with their parent strings and are then converted to hadrons
using the Lund string fragmentation model \cite{lund,jetset}
after an average proper formation time of 0.7 fm/c. Dynamics of the
resulting hadronic matter is described by a relativistic transport model
(ART) \cite{art}, which has been improved to include baryon-antibaryon 
production from meson-meson interactions and their
annihilation using cross sections given in Ref. \cite{gjwang,ko}.
Also, the $K^*$ resonances are explicitly treated by including their
production from pion-kaon and pion-rho scatterings \cite{brown}
and the inverse reactions of decay and absorption.

We first determine the parameters in the AMPT model by fitting
the experimental data from central Pb+Pb collisions at center of 
mass energy of 17 AGeV \cite{NA49}. 
Specifically, to describe the measured net baryon rapidity distribution,   
we have included in the Lund string fragmentation model the popcorn 
mechanism for baryon-antibaryon production with equal probabilities for 
baryon-meson-antibaryon and baryon-antibaryon configurations 
\cite{ampt}. Also, to account for pion and the enhanced kaon 
yields in the preliminary data from the same reaction, 
we have modified two other parameters in the Lund string fragmentation model,
following the expectation that the string tension is increased in 
the dense matter formed in the initial stage of heavy ion collisions. 

In the Lund string fragmentation model as implemented in the JETSET/PYTHIA
routine \cite{jetset}, one first assumes that a string
fragments into quark-antiquark pairs with a Gaussian distribution in 
transverse momentum. Also, a suppression factor of 0.30 is used for
strange quark-antiquark pair production relative to the light
quark-antiquark pair production. Hadrons are then formed from these 
quarks and antiquarks. For a hadron with a given transverse momentum 
$m_{\perp}$ determined by those of its quarks, its longitudinal momentum 
is given by the splitting function \cite{lund,jetset},
\begin{equation}
f(z) \propto z^{-1} (1-z)^a \exp (-b~m_{\perp}^2/z),
\label{lund}
\end{equation}
where $z$ is the light-cone momentum fraction of the produced hadron
with respect to that of the fragmenting string.
Based on the Schwinger mechanism for particle production in a strong
field, the production probability is proportional to 
$\exp (-\pi m_\perp^2/\kappa)$, where $\kappa$ is the string tension, 
i.e., energy in a unit length of string. The average squared transverse 
momentum of produced particle is thus proportional to $\kappa$. Increasing
the string tension then leads to a broader distribution of the transverse
momenta of produced quark-antiquark pairs and also a reduced suppression
for strange quark-antiquark pairs. Since the average squared
transverse momenta of produced particles obtained from Eq. (\ref{lund}) 
is $\langle p_\perp^2 \rangle=[b(2+a)]^{-1}$ for massless particles,
the two parameters $a$ and $b$ are approximately related to the
string tension by $\kappa\propto [b(2+a)]^{-1}$.  In the HIJING model, 
the default values for $a$ and $b$ are 0.5 and 0.9 GeV$^{-2}$, 
respectively.  We change their values to 2.2 and 0.5 GeV$^{-2}$
in order to increase, respectively, the pion and kaon multiplicities.
These values of $a$ and $b$ correspond to a 7\% increase of the string 
tension. 

\begin{figure}[h]
\centerline{\epsfig{file=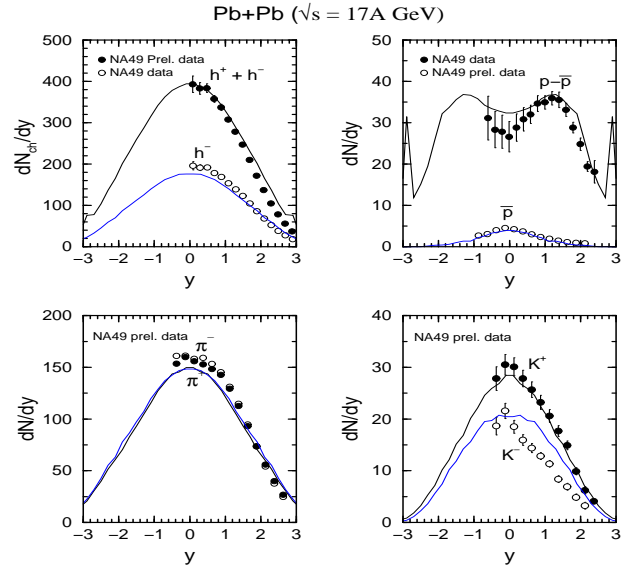,width=3.2in,height=3.2in,angle=0}}
\caption{Rapidity distributions of total and negatively charged particles 
(upper left panel), net-protons and antiprotons (upper right panel), charged 
pions (lower left panel), and charged kaons (lower right panel)
in heavy ion collisions at $\sqrt s=17$ AGeV. The circles
are the experimental data for 5\% most central Pb+Pb collision from the NA49
Collaboration, and the solid curves are the AMPT model calculations for 
impact parameters of $b\leq 3$ fm.} 
\label{fig1}
\end{figure}

Results from the AMPT model for Pb+Pb collisions at $\sqrt s=17$ AGeV,
based on the above modification of the HIJING and ART models, 
are shown in Fig. \ref{fig1}. It is seen that our model gives
a reasonable description of the data on the rapidity distributions of total 
and negatively charged particles, net-protons and antiprotons, 
charged pions, and charged kaons. Since the probability for minijet
production is very small in collisions at SPS energies, the partonic
stage does not play any role in the collisions. We find that final-state 
hadronic scatterings reduce the proton and antiproton yields, but increase 
the production of kaons and antikaons by about 20\%. 
In contrast, kaon yields in the default HIJING model are smaller than 
our final results by about 40\%.

\begin{figure}[h]
\centerline{\epsfig{file=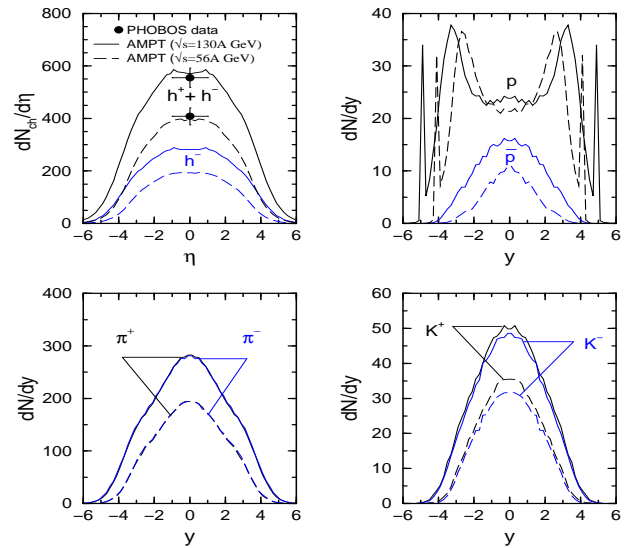,width=3.2in,height=3.25in,angle=0}}
\caption{Same as Fig. \ref{fig1} but for Au+Au collisions 
at $\sqrt s=56$ and 130 AGeV. The solid circles are the
PHOBOS data for 6\% most central collisions while the curves are the AMPT
calculations for $b\leq 3$ fm.} 
\label{fig2}
\end{figure}

Since the number of strings associated with soft interactions in the
HIJING model depends weakly on the colliding energy, the parameters
in the string fragmentation model are not expected to change much 
with the energy.  We thus use the same parameters determined from the 
experimental data at SPS to study heavy ion collisions at 
RHIC energies. In Fig. \ref{fig2}, the results for central Au+Au collisions 
at center of mass energies of 56 AGeV (dashed curves) 
and 130 AGeV (solid curves) are 
shown together with the data from the PHOBOS collaboration \cite{phobos}.
The measured total charged particle multiplicities at mid-pseudorapidity 
at both energies are well reproduced by our model. 

\begin{figure}[h]
\centerline{\epsfig{file=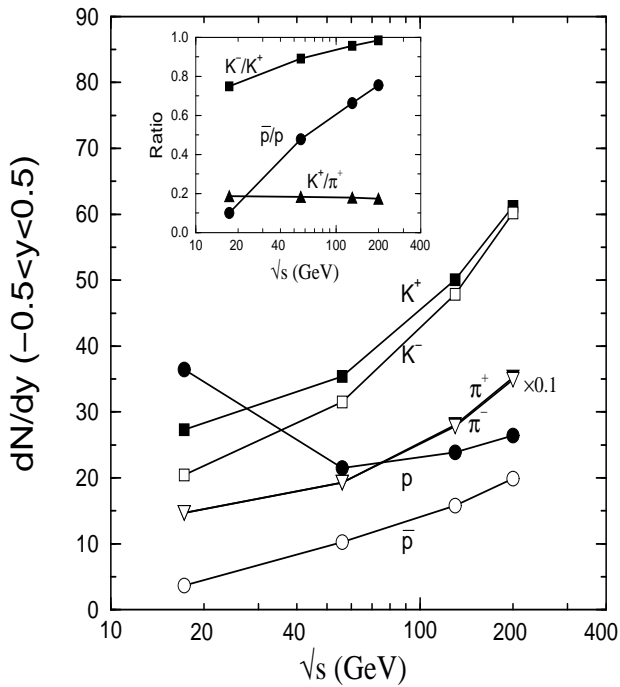,width=3.2in,height=4.5in,angle=0}}
\caption{Energy dependence of charged particle yields at mid-rapidity.   
The ratios of $K^-/K^+$, $\bar p/p$ and $K^+/\pi^+$ are 
shown in the insert.}
\label{fig3}
\end{figure}

The energy dependence of charged particle yields at mid-rapidity 
from the SPS to RHIC energies is shown in Fig. \ref{fig3}. 
The proton yield is seen to have a minimum at energies between SPS and 
the highest energy at RHIC, while antiproton yield increases almost 
linearly with $\ln s$. 
As a result, the $\bar p/p$ ratio increases rapidly from about 0.1 at SPS 
to about 0.8 at the RHIC energy of $\sqrt s=200$ AGeV, indicating 
the formation of a nearly baryon-antibaryon symmetric matter at high energies. 
Meson yields in general exhibit a faster increase with energy; 
in particular, we find that the $K^+/\pi^+$ ratio is almost constant 
within this energy range, suggesting the approximate chemical equilibrium 
for strangeness production.  The $K^-/K^+$ ratio increases gradually 
from 0.7 at SPS to about 1.0 at $\sqrt s=200$ AGeV as a result of 
the nearly baryon-antibaryon symmetric matter formed at high energies. 

\begin{figure}[h]
\centerline{\epsfig{file=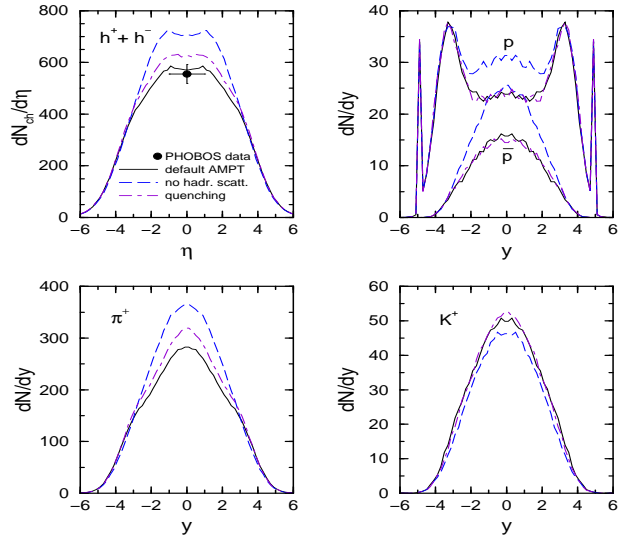,width=3.2in,height=3.2in,angle=0}}
\caption{Rapidity distributions of charged particles for central
($b\leq 3$ fm) Au+Au collisions at 130 AGeV from AMPT model
with default parameters (solid curves), without hadronic scatterings
(dashed curves), and with jet quenching of $dE/dx = 1$ GeV/fm  
(dot-dashed curves).} 
\label{fig4}
\end{figure}

To see the effects of hadronic interactions, we show in Fig. \ref{fig4}
by dashed curves the rapidity distributions of charged particles
obtained from the AMPT model without the ART model for central Au+Au
collisions at 130 AGeV. In this case, there is a significant increase in the
numbers of total charged particles, pions, protons, and antiprotons
at midrapidity. The kaon number is, on the other hand, reduced slightly. 
As a result, the ratios of $\bar p/p$ and $K^+/\pi^+$ in the absence
of final-state hadronic interactions are $0.80$ and $0.13$, respectively, 
instead of $0.66$ and $0.18$ from the default AMPT model.
We note that although the default HIJING \cite{xnwang} gives a total 
charged particle multiplicity at midrapidity that is consistent with 
the PHOBOS data, including hadronic scatterings would reduce its 
prediction appreciably.

Effects of partonic dynamics on the final hadronic observables
can also be studied in the AMPT model. Turning off the partonic cascade 
in the AMPT model, we find that this leads to less than $\sim 5\%$ 
change in the final charged particle yields at $\sqrt s=130$ AGeV. 
This indicates that the multiplicity distribution of hadrons are not 
very sensitive to parton elastic scatterings. To take into account 
the effect of parton inelastic collisions, which
are mostly responsible for energy loss, we include in the AMPT model also
the default jet quenching, i.e., an energy loss of $dE/dx = 1$ GeV/fm, 
before minijet partons enter the ZPC parton cascade. The results with jet 
quenching for central Au+Au collisions at 130 AGeV are shown in
Fig.~\ref{fig4} by the dot-dashed curves. 
We see that the quenching effects are larger for pions than 
for kaons, protons, and antiprotons. Since the present calculations from 
the AMPT model without jet quenching already reproduce 
the data at the energy of 130 AGeV, and further inclusion of jet 
quenching of $dE/dx = 1$ GeV/fm entails a 10\% increase of 
the final yield of total charged particles at midrapidity, 
our results for the rapidity distribution of charged particles are 
thus consistent with none or a weak jet quenching at this energy. 

We note that without initial nuclear shadowing on parton production 
the charged particle multiplicity at mid-rapidity at 130 AGeV 
increases by about 30\%.  This increase can nevertheless be offset by 
using different values for the parameters in the Lund string fragmentation. 
Since nuclear shadowing has negligible effects at SPS energies due to 
insignificant production of minijets, to reproduce both SPS and RHIC 
data using the same parameters requires the inclusion of nuclear shadowing 
on parton production. 

In summary, using a multiphase transport model (AMPT), which includes 
both initial partonic and final hadronic interactions, we have studied 
the rapidity distributions of charged particles such as protons,  
antiprotons, pions, and kaons in heavy ion collisions at RHIC.
With the model parameters constrained by central Pb+Pb collisions 
at $\sqrt s$ = 17 AGeV at SPS, the theoretical results 
on the total charged particle multiplicity at midrapidity in 
central Au+Au collisions at $\sqrt s$ = 56 and 130 AGeV 
agree quite well with the data from the PHOBOS collaboration.
We find that the antiproton to proton ratio at mid-rapidity increases 
appreciably with $\sqrt s$, indicating the approach to a near 
baryon-antibaryon symmetric matter in high energy collisions. 
Furthermore, the $K^+/\pi^+$ ratio is almost constant within the 
energy range studied here, suggesting the approximate chemical 
equilibrium for strangeness production in these collisions.
These hadronic observables are, however, less sensitive to 
the initial partonic interactions than the final hadronic interactions. 
To observe the effects of the partonic matter formed in the initial stage 
thus requires measurements of other observables such as $J/\psi$ 
suppression \cite{jpsi}, the elliptic flow \cite{v2}, and 
high $p_\perp$ spectra \cite{quench}.  The magnitude of elliptic flow 
has been shown to be sensitive to the parton-parton cross sections 
in the ZPC parton cascade model \cite{elliptic},  and the $J/\psi$ 
suppression results using the AMPT model indicate that the 
partonic matter plays a much stronger role than the hadronic matter 
\cite{jpsia}.

\medskip

The work of Z.L., S.P. and C.M.K. was supported by the National 
Science Foundation under Grant No. PHY-9870038, the Welch
Foundation under Grant No. A-1358, and the Texas Advanced Research
Program under Grant No. FY99-010366-0081, while that of B.A.L. 
was supported by the National Science Foundation under Grant 
No. PHY-0088934 and Arkansas Science and Technology Authority 
Grant No. 00-B-14.

\end{multicols}

\end{document}